\documentclass{ws-mpla}

\begin{document}

\markboth{C.~H. Hyun, J.~W. Shin, S. Ando}
{Parity violation in $d \vec{\gamma} \to np$
with pionless effective theory}

\catchline{}{}{}{}{}

\title{PARITY VIOLATION IN $d\vec{\gamma} \to np$
WITH PIONLESS EFFECTIVE THEORY}

\author{\footnotesize CHANG HO HYUN}

\address{Department of Physics Education, Daegu University\\
Gyeongsan 712-714, Republic of Korea, \\
hch@daegu.ac.kr}

\author{JAE WON SHIN}

\address{Department of Physics, Sungkyunkwan University\\
Suwon 440-746, Republic of Korea}

\author{SHUNG-ICHI ANDO}

\address{Theoretical Physics Group, School of Physics and Astronomy,
The University of Manchester, \\
Manchester M13 9PL, United Kingdom}

\maketitle

\pub{Received (Day Month Year)}{Revised (Day Month Year)}

\begin{abstract}
We consider a pionless effective theory with dibaryon fields for 
the description of the weak process involving two nucleons.
We construct leading order Lagrangians that contain
nucleon-dibaryon weak coupling constants.
We calculate the physical observable in the photodisintegration 
of the deuteron at threshold and obtain the result in terms of 
the nucleon-dibaryon weak coupling constants. 
Relation to existing calculations is discussed.

\end{abstract}

\ccode{PACS Nos.: 21.30.Fe}

\section{Introduction}	

\indent
It is quite recent that the effective field theory (EFT)
has put its first step toward the realm of hadronic weak 
interaction. \cite{hyun01,zhu05}
An experiment to measure parity-violation (PV)
effects in polarized neutron capture by a proton at LANSCE
and its upgrade at SNS have evoked interest in exploring
the PV phenomena in few nucleon systems in the context
of EFT. \cite{hyun07,liu07,desplanques08}
The results obtained thus far indicate that EFT is a working
machine for the study of the PV processes; 
perturbative expansion converges 
well in both interactions and transition operators.
Inspired by the precedent success of the EFT in PV,
we try to formulate the PV EFT as simple as possible so that
it will be feasible to explore various few nucleon PV phenomena 
with it.
Pionless EFT with dibaryon fields has proven one of the most 
efficient EFTs applicable to few nucleon systems at low energies.
\cite{ando05,ando06,ando08}
PV EFT with dibaryon fields was already considered by Savage 
\cite{savage01}, but only a part of more general pionless PV 
Lagrangian was accounted in the work.
In this work, we construct pionless PV Lagrangian with dibaryon
fields at leading order, and calculate the PV observable in
$d\vec{\gamma} \rightarrow n p$ at threshold. 
We discuss how the present result can be
compared with previous theoretical calculations.

\section{Theory}

\indent
PV dibaryon-nucleon-nucleon ($dNN$) Lagrangian 
can be written as
\begin{eqnarray}
{\cal L}^{dNN}_{\mbox{\tiny PV}} = \sum_{\Delta T} 
{\cal L}^{\Delta T}_{\mbox{\tiny PV}}
\end{eqnarray}
where $\Delta T$ is the isospin change at the $dNN$ vertex.
For the two nucleon system, 
\begin{eqnarray}
{\cal L}^{dNN}_{\mbox{\tiny PV}} = 
{\cal L}^0_{\mbox{\tiny PV}} + {\cal L}^1_{\mbox{\tiny PV}}.
\end{eqnarray}
Because of the total angular momentum conservation,
allowed transitions due to the PV interaction
for the lowest angular momentum states are
$^1 S_0 \leftrightarrow {}^3 P_0$, and $^3 S_1 \leftrightarrow {}^1 P_1$
due to ${\cal L}^0_{\mbox{\tiny PV}}$, and $^3 S_1 \leftrightarrow {}^3 P_1$
due to ${\cal L}^1_{\mbox{\tiny PV}}$.
Introducing dimensionless PV $dNN$ coupling constants
$h^{\Delta T}_{\mbox{\tiny dNN}}$, we have the PV $dNN$ Lagrangians
\begin{eqnarray}
{\cal L}^0_{\mbox{\tiny PV}} &=&
\frac{h^{0 s}_{\mbox{\tiny dNN}}}{2 \sqrt{2} \rho_d\, m_N^{5/2}} 
s^\dagger_a\, N^T 
\sigma_2 \sigma_i  \tau_2 \tau_a 
\frac{i}{2} \left(\stackrel{\leftarrow}\nabla - 
\stackrel{\rightarrow}\nabla \right)_i N  \label{eq:Lwk0s}\\
& & + 
\frac{h^{0 t}_{\mbox{\tiny dNN}}}{2 \sqrt{2} \rho_d\, m_N^{5/2}} \,
t^\dagger_i\, N^T \sigma_2 \tau_2 
\frac{i}{2} \left(\stackrel{\leftarrow}\nabla - 
\stackrel{\rightarrow}\nabla \right)_i N
+{\rm h.c}, \label{eq:Lwk0t}
\end{eqnarray}
for the isospin conserving part and
\begin{eqnarray}
{\cal L}^1_{\mbox{\tiny PV}} =
i \frac{h^1_{\mbox{\tiny dNN}}}{2 \sqrt{2} \rho_d\, m_N^{5/2}} \,
\epsilon_{ijk}\, t^\dagger_i\, N^T \sigma_2 \sigma_j \tau_2  \tau_3
\frac{i}{2} \left(\stackrel{\leftarrow}\nabla - 
\stackrel{\rightarrow}\nabla \right)_k N
+{\rm h.c} \label{eq:Lwe1t}
\end{eqnarray}
for the isospin changing one.
$s_a$ and $t_i$ are the two-nucleon dibaryon fields in $^1 S_0$ and
$^3 S_1$ states, respectively.
$\sigma$'s and $\tau$'s 
in Eqs.~(\ref{eq:Lwk0s},\ref{eq:Lwk0t},\ref{eq:Lwe1t}) project the
neutron-proton system into $^3 P_0  $, $^1 P_1$ and $^3 P_1$ states, 
respectively.
%


\section{Diagrams and Results}

\begin{figure}[tbp]
\centerline{\psfig{file=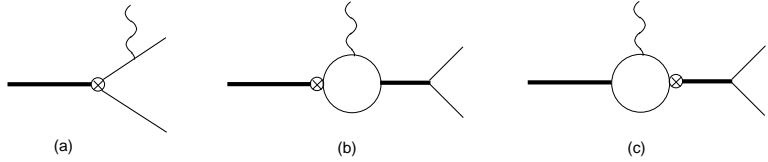,width=4.0in}}
\vspace*{8pt}
\caption{Leading diagrams for $d\vec{\gamma} \to np$  process.
Thick line represents the {\it dressed} dibaryon fields, thin line
the nucleon and the wavy line the photon. 
Cross circle stands for PV $dNN$ vertex, and the $dNN$ vertices
without any mark for the PC interaction.
\protect\label{fig:dgnp}}
\end{figure}

An observable is the PV photon polarization defined as
\begin{eqnarray}
P_\gamma = \frac{\sigma_+ - \sigma_-}{\sigma_+ + \sigma_-}
\end{eqnarray}
where $\sigma_{+(-)}$ is the total cross section with the photon 
of helicity $+1$ $(-1)$.
With PV interactions, $P_\gamma$ is given as \cite{holstein05}
\begin{eqnarray}
P_\gamma = - 2 \frac{{\cal M}_{\tilde{E1}}}{{\cal M}_{M1}},
\end{eqnarray}
where ${\cal M}_{M1}$ and ${\cal M}_{\tilde{E1}}$ are the 
parity-conserving (PC) M1 and parity-violating E1 amplitudes,
respectively. PC M1 amplitude has been calculated very 
accurately in the pionless EFT with dibaryons \cite{ando05,ando06},
and we employ the result in this work.

Feynman diagrams for the process are depicted in Fig.~\ref{fig:dgnp}.
It has been shown in the literature that the PV polarization 
depends on isoscalar and isotensor components of the meson-exchange
PV potential.
In our framework, isospin conserving weak Lagrangians given
by Eqs.~(\ref{eq:Lwk0s},\ref{eq:Lwk0t}) contribute to ${\cal L}_{\tilde{E1}}$.
In the diagrams (a) and (b), we have the initial $^3 S_1$ state, and
$^1 \tilde{P}_1$ state is admixed in the initial state by the
weak Lagrangian ${\cal L}^{0t}_{\mbox{\tiny PV}}$,
$
| ^1 \tilde{P}_1 \rangle \propto 
{\cal L}^{0t}_{\mbox{\tiny PV}} | ^3 S_1 \rangle \propto h^{0t}_{\mbox{\tiny dNN}}.
$
%
Since the leading E1 transition operator conserves the total spin,
we have the PV amplitude from the diagrams (a) and (b) as
\begin{eqnarray}
{\cal M}_{\rm a,b} = \langle ^1 S_0 | \hat{E}_1 |^1 \tilde{P}_1 \rangle
\propto h^{0t}_{\mbox{\tiny dNN}},
\end{eqnarray}
where $\hat{E}_1$ is the E1 transition operator.
For the diagram (c) we have a parity admixture in the final state,
$
| ^3 \tilde{P}_0 \rangle \propto {\cal L}^{0s}_{\mbox{\tiny PV}}
| ^1 S_0 \rangle \propto h^{0s}_{\mbox{\tiny dNN}}.
$
%
We then have the PV transition amplitudes
\begin{eqnarray}
{\cal M}_{\rm c} = \langle ^3 \tilde{P}_0 | \hat{E}_1 | ^3 S_1 \rangle
\propto h^{0s}_{\mbox{\tiny dNN}},
\end{eqnarray}

We employ the convection current operator 
of the nucleon for $\hat{E}_1$, which is proportional to the 
nucleon momentum. Counting the order of on-shell momentum in each
diagram, diagram (a) is suppressed to diagrams (b) and (c) by $p^2$ 
where $p$ is the outgoing nucleon momentum.
Consequently diagrams (b) and (c) give leading contributions,
and we obtain the results in the $p\to 0$ limit as
\begin{eqnarray}
i {\cal M}_{\rm b} = 
\frac{a_0\, \gamma h^{0t}_{\mbox{\tiny dNN}}}{6 \rho_d m^{5/2}_N}
\sqrt{\frac{\gamma \rho_d}{1-\gamma \rho_d}} , \,\,\,\,\,\,
i {\cal M}_{\rm c} =
\frac{a_0\, \gamma h^{0s}_{\mbox{\tiny dNN}}}{6 \rho_d m^{5/2}_N}
\sqrt{\frac{\gamma \rho_d}{1-\gamma \rho_d}}.
\label{eq:Mbc}
\end{eqnarray}
$a_0$ is the scattering length for the $^1 S_0$ scattering state, 
$\rho_d$ is the effective range for the $^3 S_1$ bound state,
and $\gamma = \sqrt{m_N\, B}$, 
where $B$ is the deuteron binding energy.
With the PC M1 amplitude obtained from the same theory, 
we have 
\begin{eqnarray}
P_\gamma = - \frac{a_0 \gamma}{6 \sqrt{2 \pi m_N \rho_d}}
\frac{h^{0s}_{\mbox{\tiny dNN}} + h^{0t}_{\mbox{\tiny dNN}}}{
\kappa_1 ( 1 - a_0 \gamma) - \frac{L_1}{2}a_0 \gamma^2},
\end{eqnarray}
where $\kappa_1 = \mu_V/2 = 2.35$ and $L_1$ is a low energy constant
fixed from the $np \to d\gamma$ total cross section at threshold.

\section{Discussion}

Using the pionless EFT with dibaryon fields,
we calculated the PV polarization in $d\vec{\gamma} \to np$ at
threshold.
Though the use of the EFT makes the calculation simple 
and easy, physical meaning of the result is not as 
transparent as the one obtained from the meson exchange
picture of the weak $NN$ potential, 
so called the DDH potential \cite{ddh80}.
In order to better figure out the physical meaning of the
present result, and eventually make the EFT a useful
method for PV phenomenology, it is necessary to have the
relation between the ``new" and ``old" languages.
There are several steps to establish the connection.
First, we have to relate the weak meson-nucleon coupling
constants in the DDH potential to the PV $dNN$ vertices. 
Starting point may be the pionful EFT. Expanding the 
pionful PV potential to a certain order, and then integrating
out the pion degree of freedom, we obtain the relation
between pionful and pionless theories.
Second, we have to be cautious in treating the EM operators
for the external photons. 
Most of the calculation in the literature assumes Siegert's
theorem for the E1 operator.
In Ref.~\refcite{hyun01}, it was shown that the PV asymmetry
in $\vec{n}p \to d\gamma$ with convection current (adopted
in this work) overestimates the result with Siegert's theorem
by a factor of 2.9. It was shown in the same reference that
the meson-exchange terms strongly cancel the convection term
contribution, and thus make the net result similar to the
one with Siegert's theorem.
Understanding the role of higher order operators may be crucial
to put a bridge between pionful and pionless results.
If the relation is well understood and thus the pionless
EFT with dibaryon fields becomes an understandable language
in the hadronic weak interaction, the method can be applied
to a broader range of few-body PV phenomena. 

\section*{Acknowledgments}

Work of CHH was supported by the Daegu University Research Grant, 2008.
Work of JWS was supported in part by the Korea Science and Engineering
Foundation grant funded by the Korean Government (MOST)
(No.M20608520001-07B0852-00110).
SA is supported by STFC grant number PP/F000448/1.



\begin{thebibliography}{99}
\bibitem{hyun01} C. H. Hyun, T.-S. Park and D.-P. Min, 
{\it Phys. Lett. B} {\bf 516}, 321 (2001).

\bibitem{zhu05} S.-L. Zhu {\it et al.}, 
{\it Nucl. Phys. A} {\bf 748}, 435 (2005).

\bibitem{hyun07} C. H. Hyun, S. Ando and B. Desplanques,
{\it Phys. Lett. B} {\bf 651}, 257 (2007); 
{\it Eur. Phys. J. A} {\bf 32}, 513 (2007).

\bibitem{liu07} C.-P. Liu,
{\it Phys. Rev. C} {\bf 75}, 065501 (2007).

\bibitem{desplanques08} B. Deslanques, C. H. Hyun, S. Ando and C.-P. Liu,
{\it Phys. Rev. C} {\bf 77}, 064002 (2008).

\bibitem{ando05} S. Ando and C. H. Hyun,
{\it Phys. Rev C} {\bf 72}, 014008 (2005).

\bibitem{ando06} S. Ando, R. H. Cyburt, S. W. Hong and C. H. Hyun,
{\it Phys. Rev. C} {\bf 74}, 025809 (2006).

\bibitem{ando08}
S. Ando {\it et al.}, 
{\it Phys. Lett. B} {\bf 668}, 187 (2008).

\bibitem{savage01} M. J. Savage, {\it Nucl. Phys. A} {\bf 695}, 365 (2001).

\bibitem{holstein05} B. R. Holstein, {\it Fizika B} {\bf 14}, 165 (2005).

\bibitem{ddh80} B. Desplanques, J. F. Donoghue and B. R. Holstein,
{\it Ann. Phys. (N.Y.)} {\bf 124}, 449 (1980).

\end{thebibliography}
\end{document}